\documentclass[%
 reprint,
 amsmath,amssymb,
 aps,
]{revtex4-2}

\usepackage{multirow}
\usepackage{graphicx}% Include figure files
\usepackage{dcolumn}% Align table columns on decimal point
\usepackage{bm}% bold math
\usepackage{hyperref}% add hypertext capabilities
\usepackage{siunitx}
\newcommand{\ssb}[1]{_{\textrm{\tiny{#1}}}}
\newcommand{\spr}[1]{^{\textrm{\tiny{#1}}}}

\begin{document}

\preprint{APS/123-QED}

\title{A search for neutron-to-hidden-neutron oscillations in a ultra-cold neutron beam}

\author{G. Ban}
\author{J. Chen}
\author{T. Lefort}
\author{O. Naviliat-Cuncic}
 \altaffiliation[Also at ]{Michigan State University, Michigan, United States.}
\author{W. Saenz-Arevalo}
\email{william.saenz@lpnhe.in2p3.fr}
\altaffiliation[Present address: ]{Laboratoire de Physique Nucléaire et des Hautes Énergies, Paris, France}
\affiliation{Université de Caen Normandie, ENSICAEN, CNRS/IN2P3, LPC Caen UMR6534, F-14000 Caen, France.}

\author{P.-J. Chiu}
\affiliation{University of Zurich, Zurich, Switzerland.}

\author{B. Clément}
\author{P. Larue}
\author{G. Pignol}
\author{S. Roccia}
\email{roccia@lpsc.in2p3.fr}
 \altaffiliation[Also at ]{Institut Laue-Langevin, Grenoble, France.}
\affiliation{Université Grenoble Alpes, CNRS/IN2P3, Grenoble INP, Laboratoire de Physique Subatomique et de Cosmologie (LPSC), 38000  Grenoble, France.}

\author{M. Guigue}
\affiliation{Sorbonne Université, Université Paris Cité, CNRS/IN2P3, Laboratoire de Physique Nucléaire et de Hautes Énergies (LPNHE), 75005 Paris, France.}

\author{T. Jenke}
\author{B. Perriolat}
\affiliation{Institut Laue-Langevin, 71 avenue des Martyrs, CS 20156, 38042 Grenoble Cedex 9, France.}

\author{P. Schmidt-Wellenburg}
\affiliation{Paul Scherrer Institute, Villigen, Switzerland.}

\date{\today}

\begin{abstract}

Models that postulate the existence of hidden sectors address contemporary questions, such as the source of baryogenesis and the nature of dark matter. Among the possible mixing processes, neutron-to-hidden-neutron oscillations have been repeatedly tested with ultra-cold neutron storage and passing-through-wall experiments in the range of small ($\delta m<2$ peV) and large mass splitting ($\delta m>10$ neV), respectively. In this work, we present a new constraint in the oscillation parameter space derived from neutron disappearance in ultra-cold neutron beam experiments. The overall limit, which covers the intermediate mass-splitting range, is given by $\tau_{nn'}> 1 \textrm{\,s for\,\,} |\delta m| \in [2,69] \,\textrm{peV}  \textrm{  (95\% C.L.)}$.

\end{abstract}


\maketitle

Hidden sectors have been proposed in several contexts. Initially, Lee and Yang postulated the existence of `mirror particles' to explain parity symmetry breaking in the weak interaction \cite{LeeYang1956}. More than 20 years after the first experiments that evidenced P and CP breaking \cite{Wu,Christenson}, R. Foot proposed the existence of a mirror sector that hosts copies of all known particles and their interactions \cite{Foot_1992}. It was pointed out that this extra sector might potentially mix with ordinary particles through non-Standard-Model (SM) interactions, besides gravity. In recent years, the double-degenerated mirror theory of hidden particles has been formulated as a particular case of more extensive models. For example, Dvali and Redi \cite{Dvali_2009} explained that in order to solve the hierarchy problem between the scales of weak nuclear forces and gravity, the number of allowed SM copies can go up to $10^{32}$. Also, from a geometrical point of view, the disappearance of particles could correspond to ordinary particles transitioning into different layers (branes) of a high-dimensional bulk \cite{Sarrazin_2012}. The universe we observe today might represent a three-dimensional sheet embedded in hyperspace, where multiple versions of SM particles are constrained to live. This serves to present hidden matter as a candidate for dark matter.

For all the aforementioned models, the mixing of neutral matter with neutral hidden matter is described by the same phenomenology \cite{Sarrazin_2010}. In particular, if the two-sector model is adopted, the Hamiltonian describing the oscillation of neutrons ($n$) into hidden neutrons ($n'$) is:
\begin{equation}
\mathcal{\hat{H}}_{nn'}= \begin{pmatrix}
E_n & \epsilon_{nn'} \\
\epsilon_{nn'} & E_{n'}
\end{pmatrix}=\begin{pmatrix}
m_n + \Delta E & \epsilon_{nn'} \\
\epsilon_{nn'} & m_n + \delta m
\end{pmatrix},
\label{eq:first_h}
\end{equation}
where $E_n$ ($E_{n'}$) is the neutron (hidden neutron) total energy and $\epsilon_{nn'}=\tau_{nn'}^{-1}$ is the mass mixing parameter (with $\hbar=c=1$). In Eq.\ (\ref{eq:first_h}), $E_n$ has been separated into the neutron rest energy $m_n$ and its energy due to interactions $\Delta E$. $E_{n'}$ is expressed as a function of the mass splitting $\delta m= m_{n'}-m_n + V'$, which implicitly contains the spin-independent hidden-neutron interactions. A similar, but not equivalent, description stems from considering hidden spin-dependent interactions, such as scenarios with hidden (mirror) magnetic fields \cite{berezhiani_2009}.

In the past, $n-n'$ oscillations have been extensively studied in ultra-cold neutron (UCN) experiments by observing at the neutron disappearance in storage bottles \cite{Ban_2007,SEREBROV2008181,Altarev_2009,Berezhiani_2018_New,psi2021}, and in high flux neutron setups by monitoring neutron regeneration behind dense neutron stoppers \cite{Sarrazin_2015_probing,Stasser_2021,stereo,Broussard_2022}. Since most of the highly sensitive $n-n'$ measurements at low $\delta m$ did not find any significant signal, there is great interest in improving the sensitivity at larger mass splittings: $\delta m>1$ peV. This work presents the results of the first experiment probing $n-n'$ oscillations for $\delta m \in [3,66]$ peV via neutron disappearance in UCN beams. The search was conducted by lifting the neutron-hidden neutron energy degeneracy ($\Delta_{nn'}=\Delta E-\delta m$) by applying an external magnetic field: $\Delta E= \mu_n B$, where $\mu_n$ is the neutron magnetic moment. If the neutron energy matches the mass splitting ($\Delta_{nn'}=0$), oscillations would occur with maximum amplitude.

\begin{figure}[htb]
\includegraphics[scale=0.468]{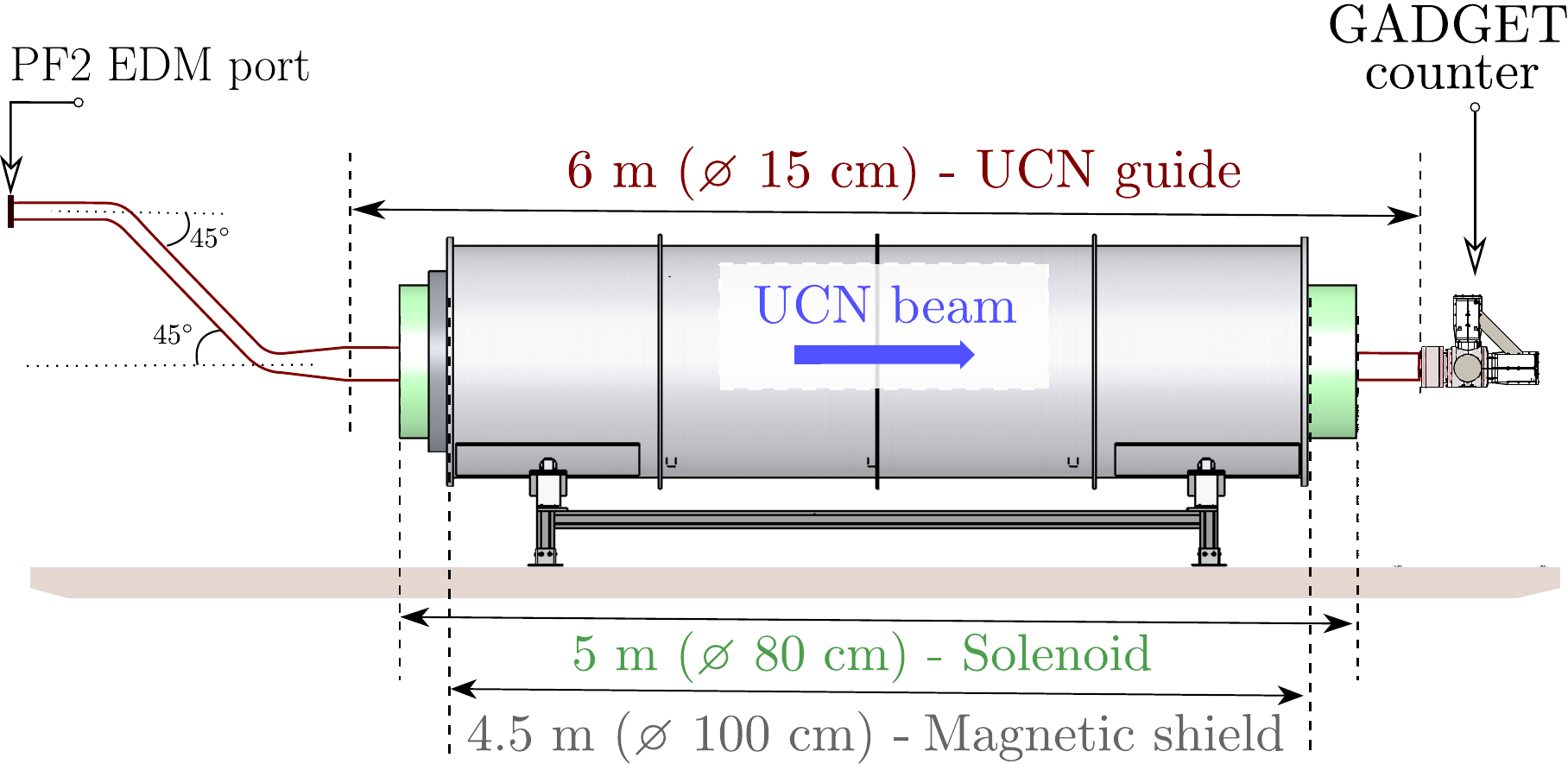}
    \caption{Schematic side view of the experimental setup.}
    \label{fig:setup}
\end{figure}

The experiment took place in autumn 2020 at the high flux reactor of the Institut Laue Langevin (ILL) in Grenoble, France. The UCN beam, originating from the ILL's neutron turbine, was transported from the EDM port at the PF2 experimental hall through a vacuumed NiMo-coated guide to the detector. As the UCN counting rate at the EDM port can be as high as $\sim 500000$ counts/s, the experiment was conducted using the fast and stable UCN gaseous detector GADGET \cite{Ayres_2021_n2EDM}. The setup featured a 6-m-long horizontal UCN guide, on top of which the magnetic field generated from a 5-m-long solenoid was applied (see Fig.\ \ref{fig:setup}). This field was shaped and screened from external sources by a cylindrical mu-metal shield. Due to the constant-power operation mode of ILL's reactor, neutron disappearance arising from $n-n'$ oscillations during the bouncing trajectories of UCN in this setup would cause a drop in the neutron counts at the GADGET detector.

\begin{figure*}
\includegraphics[scale=0.68]{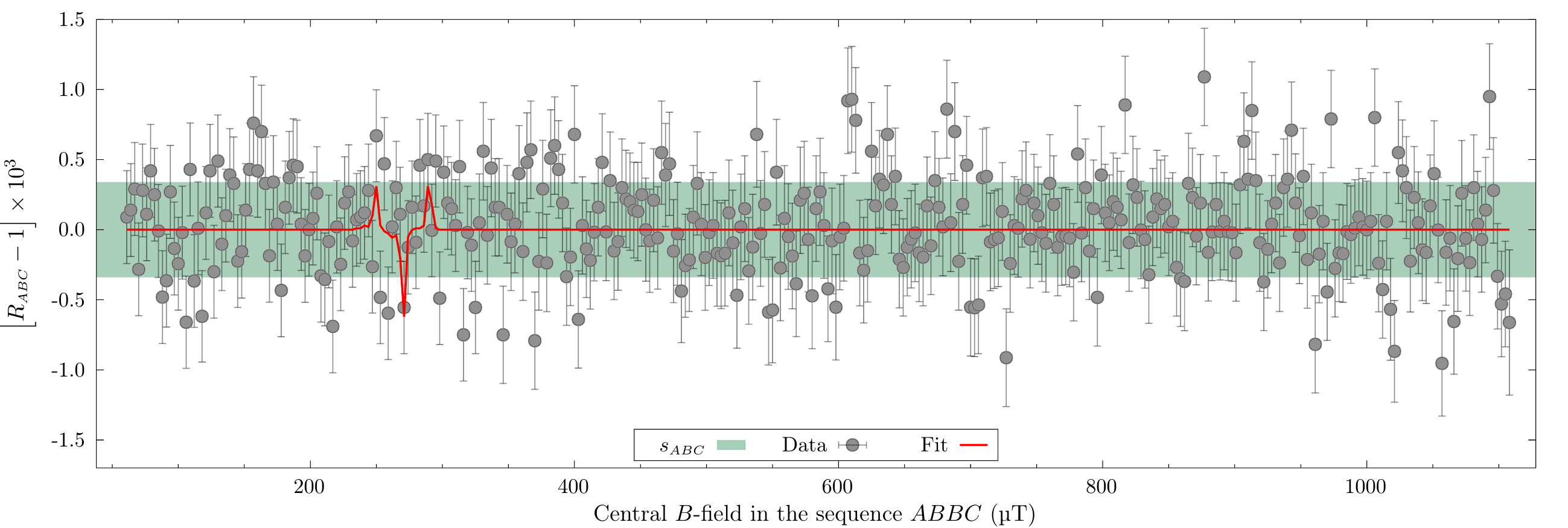}
\caption{Normalized UCN flux $R_{ABC}$ as a function of the applied cycle central field $B$. The data points correspond to the weighted average over the 14 scans, independently of the applied field direction. The displayed fit does not correspond to significant signal, as it slightly overcomes the $1\sigma$ band ($s\ssb{ABC}$).}
\label{fig:fit}
\end{figure*}

The study of oscillations was carried out by scanning the magnetic field in the range of $[50,1100]$ \SI{}{\micro\tesla}, i.e., $\delta m \in [3,66]$ peV. A self-normalized counting rate was measured by introducing a non-linear sequence in the $B$-field scanning process. For this purpose, each UCN delivery cycle of 200 s was divided into four windows of equal duration (44 s), in which three magnetic fields were applied as $\{A,B,B,C\}= \{B-20,\textrm{\SI{}{\micro\tesla}},B,B,B+20,\textrm{\SI{}{\micro\tesla}\}}$. The in-cycle field step of \SI{20}{\micro\tesla} was chosen larger than the resonance FWHM (\SI{1}{\micro\tesla}) so that $n-n'$ oscillations could occur only at one of the cycle magnetic field values ($A,B,$ or $C$). In this way, using the total UCN counts at fields $B$ ($N_B+N_B$), $A$ ($N_A$), and $C$ ($N_C$), the self-normalized UCN flux was conventionally defined as
\begin{equation}
R_{ABC}=\frac{N_B+N_B}{N_A+N_C} \left\{\begin{array}{ll}
= 1, & \textrm{if no oscillations} \\
< 1, & \textrm{if } B\approx \delta m/\mu_n \\
> 1, & \textrm{if } A \textrm{ or } C\approx \delta m/\mu_n.
\end{array}\right.
\label{eq:R}
\end{equation}
While 44 s were dedicated to integrate the detected UCN flux at each field value, the remaining 24 seconds were used to wait for the beam stabilization and to ramp the magnetic fields. Note that the $R_{ABC}$ observable is not only independent of the long-term variations of UCN flux (from one cycle to another) but also of linear drifts within cycles. The cycle-to-cycle magnetic-field step size was set to \SI{3}{\micro\tesla} to ensure a $n-n'$ probe of all values of $B\in[50,1100]$ \SI{}{\micro\tesla} with a resolution of \SI{1}{\micro\tesla}. In total, 3794 cycles were recorded with positive (2304) and negative (1490) field configurations during an experimental campaign of 25 days. The data-set \cite{data} consisted of 14 scans, most of which completed the entire sweep of the $B$-field range (see Fig.\ \ref{fig:fit}). 

Oscillations of neutrons into hidden neutrons would occur inside the main guide segment between two UCN wall collisions. While hidden neutrons would pass through the guide walls or the detector, ordinary neutrons are reflected along the guide and finally counted by the detector. In the latter case, the neutron wave function collapses into a pure ordinary neutron state at each wall collision, making the oscillation probability ($P_{nn'}$) reset to zero in a Zeno-like effect. The sensitivity of this experiment is governed by the magnitude of the time between wall collisions (free-flight time, $t_f$), and the total number of wall collisions ($n\ssb{coll}$). This can be seen from the neutron-oscillation probability, which, at the exact energy degeneracy lifting $\Delta_{nn'}=0$, is given by $P_{nn'}(t_f) = (t_f/\tau_{nn'})^2$. The average value of both quantities, $\bar{t}_f$ and $\bar{n}\ssb{coll}$, was estimated using Monte Carlo (MC) simulations of the UCN tracks \cite{starucn} for the current geometry. Among the simulation input parameters were the probability of diffusive reflections (1\%), the absorption coefficient upon reflection ($5.4\times 10^{-4}$), and the spectrum of initial velocities (later discussed). The simulation output yielded $\bar{t}_f=32.2$ ms and $\bar{n}\ssb{coll}=26$, which determine a rough sensitivity of $\tau_{nn'}\sim 1$ s for the entire range of scanned $B$-fields \cite{saenz_moriond}.

The search for oscillation signals with the ratio $R_{ABC}$ in Eq.\ (\ref{eq:R}) was performed by comparing it to the model prediction  
\begin{equation}
    R_{ABC}^{\textrm{theo}} = \frac{1- \langle P_{nn',B} \rangle }{1 - (\langle P_{nn',A} \rangle  +  \langle P_{nn',C} \rangle )/2 }.
    \label{eq:prediction}
\end{equation}
Here, $\langle P_{nn',B_i} \rangle$ represents the average $n-n'$ oscillation probability per UCN at magnetic field $B_i=A,B$ or $C$. These average probabilities depend on the model parameters $\tau_{nn'}$ and $\delta m$. They account for the total probability of a UCN traveling from the EDM port to the detector to oscillate into a hidden neutron. In general, for a UCN describing $n\ssb{coll}$ wall collisions, the probability of oscillating is expressed as
\begin{align}
    P_{nn'} &= P_1 + (1-P_1)P_2 + \dots + \left(\prod\limits_{i=2}^{n\ssb{coll}} (1 -P_{i-1})\right)P_{n\ssb{coll}}, \nonumber \\
    & \approx P_1 + P_2 + P_3 + \dots + P_{n\ssb{coll}},\label{eq:steps} 
\end{align}
where $P_i$ represents the oscillation probability between the $(i-1)$-th and $i$-th wall collisions of the UCN. The non-linear terms in the last equation are neglected since even for perfect energy degeneracy lifting, the oscillation probability is at most $P_{nn'} \sim 10^{-4}$ (with $\tau_{nn'}=1$ s).

Computation of $P_{nn'}(t)$ is straightforward from the analytical solution of $\mathcal{\hat{H}}_{nn'}$ for perfectly uniform magnetic fields. However, for the present work, we took into account the magnetic field gradients experienced by UCN while crossing the solenoid. This was done by implementing a numerical solution of the Liouville-Neumann equation %\cite{Liouville}
\begin{equation}
    \partial_t\hat{\rho} = -i[\hat{\mathcal{H}}_{nn'},\hat{\rho}] = -i\hat{\mathcal{H}}_{nn'}\hat{\rho}+i\hat{\rho}\hat{\mathcal{H}}_{nn'}^{\dagger},
    \label{eq:liouville}
\end{equation}
where $\hat{\rho}$ is the $2\times2$ density matrix in the basis $(\psi_n,\psi_{n'})$ representing the quantum state composed of neutrons and hidden neutrons. The magnetic field spatial description $\vec{B}(\vec{r})$ was included within the definition of $\mathcal{\hat{H}}_{nn'}$.

The average oscillation probability per neutron was determined from the solution of the Liouville-Neumann equation along multiple simulated UCN tracks. First, the velocity input parameters (magnitude and direction) of the MC simulation were optimized to reproduce a UCN time-of-flight (TOF) measurement performed at 1 m from the EDM beam port. Then, the coordinates of the wall collisions and the free-flight times were determined from the MC simulation. They were fed into the numerical solution of Eq.\ \ref{eq:liouville} where the wave function collapse at the wall collisions was manually introduced by making $(\hat\rho)_{ij}=1$ for $i=j=1$ and 0 elsewhere. Since the time evolution of $P_{nn'}(t)$ is fixated by the magnitude of $(\hat\rho)_{22}(t)$, the average oscillation was calculated as
\begin{equation}
    \langle P_{nn'}  \rangle=\frac{1}{N\ssb{UCN}}\sum\limits_{i=1}^{N\ssb{UCN}} \sum\limits_{j=1}^{n\ssb{coll$,i$}}  (\hat\rho)_{22}(t_{i,j}\spr{coll}),
\end{equation}
with $t_{i,j}\spr{coll}$ being the $j$-th collision time of the $i$-th UCN track, and $N\ssb{UCN}$ the total number of UCN tracks.

The computation of $\langle P_{nn'} \rangle$ ignored the tracks of UCN lost through the known processes: $\beta$-decay, absorption, up-scattering, and transmission at the guide walls, as by definition, they cannot be detected. $100$ tracks ($\sim$2600 free-flight steps) of UCN reaching the detector entrance window were used to compute the average probabilities as a function of $\delta m$ for $\tau_{nn'}=1$ s. This number of tracks represented a good compromise between the uncertainty on the average probability associated to its rate of converge and the total computation time. A study considering $5\times10^4$ UCN estimated this uncertainty to be within 5\% for the employed numerical algorithm time step: $\Delta t = 10$ \SI{}{\micro\second}. The average probability was then calculated for every applied $B$-field, with 500 values of $\delta m$ comprised in the range of $[0.1,1300]\,$\SI{}{\micro\tesla}$/\mu_n$. The distribution of $\delta m$ points along this range was not uniform but rather compressed around $\Delta_{nn'}=0$ to better describe the oscillation resonance behavior. The evaluation at different oscillation times was obtained from the 1 s estimation as
\begin{equation}
\langle P_{nn',B} (\delta m, \tau_{nn'})\rangle = 
\left(\frac{1\,\rm{s}}{\tau_{nn'}}\right)^2 \langle P_{nn',B}(\delta m,1\,\textrm{s}) \rangle.
\label{eq:scale}
\end{equation}
This last expression is derived from the analytical solution of $P_{nn'}$ \cite{Berezhiani_2018_New}, where the factor $\tau_{nn'}^{-2}$ can be factorized from the term depending on the energy degeneracy $\Delta_{nn'}$.

The magnetic field experienced by UCN $\vec{B}(\vec{r}\ssb{UCN})$ was estimated from $B$-field measurements along the solenoid axis $\vec{B}(0,z)$. For off-axis ($r_i>0$) estimations, the field chart $\vec{B}(0,z)$ was scaled by adding a calibrated COMSOL simulated magnetic field map. Then, using the simulated UCN track coordinates $(r_i,z_i)$, the actual field profiles were determined while assuming azimuthal symmetry, with $z$ along the solenoid axis. An analysis considering $10^3$ MC trajectories revealed deviations of the field profiles of no more than 2\% from $\vec{B}(0,z)$. This justified the neglect of radial gradients ($\nabla_rB$) of the magnetic field. Likewise, the radial component $B_r$ was observed to represent, on average, a 1\% or less of the total field magnitude. These two features allowed approximating all UCN field profiles as $\vec{B}(r_i,z_i)\approx B_z(0,z_i) \hat{z}$. Finally, since oscillations are excluded far from the solenoid with time constants of up to a few seconds \cite{Abel2020}, the average oscillation probability only considered UCN trajectory steps contained inside the solenoid volume for which $B>20\,\textrm{\SI{}{\micro\tesla}}$. 

Given that the UCN beam was not polarized, no discrimination between positive and negative field configurations could be made in the analysis. One can see that if assuming $\delta m>0$ ($<0$), spin-up neutrons would fulfill the resonance condition at $+B$ ($-B$), whereas spin-down neutrons would do it at $-B$ ($+B$). As a result, the same number of disappearing UCN is expected for both field directions. The final data-set of $R_{ABC}$ points was computed as the weighted average over all the measured scans, regardless of their field direction. 

The initial poissonian error bars on the neutron counts were scaled by a factor $s=2.23$ in order to account for the reactor power fluctuations \cite{saenz_phd}. Then, by using Eqs.\ \ref{eq:prediction} and \ref{eq:scale}, the $R_{ABC}$ data points were fitted with free parameters $\delta m$ and $\tau_{nn'}$ as shown in Fig.\ \ref{fig:fit}. The displayed fit corresponds to the best estimation $(\delta m/\mu_n, \tau_{nn'})=(271.1 \textrm{ \SI{}{\micro\tesla}},3.2 \textrm{ s})$ whose $\chi^2$ per degree of freedom is $\chi^2/\rm NDF=343.9/348$. The triple peak signature of the fit is a characteristic of the $R_{ABC}$ ratio, as it depends simultaneously on $B$, $B+20$ \SI{}{\micro\tesla}, and $B-20$ \SI{}{\micro\tesla}. Note that the fitted signal is largely contained within the $1\sigma$ band ($s_{ABC}$) defined by the data-set dispersion. The fit is not more significant than the one obtained from the null-hypothesis, computed from Eq.\ \ref{eq:prediction} with $\tau_{nn'}\rightarrow\infty$, for which $\chi^2\ssb{null}/\rm NDF=348.5/349$.  

\begin{figure}[htb]
\includegraphics[scale=0.71]{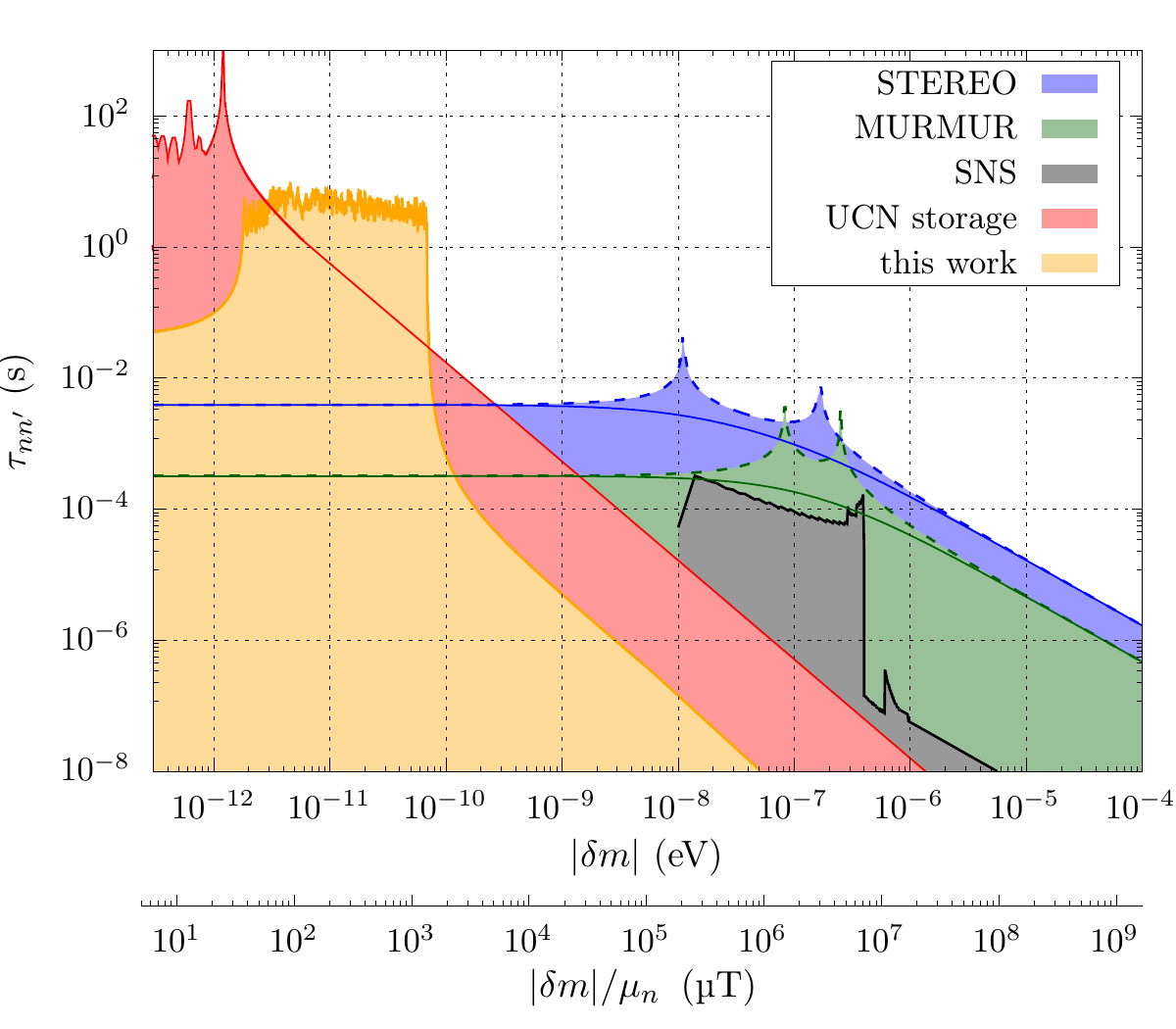}
    \caption{Exclusion of the $n-n'$ parameter space including all experimental results up to August 2022.}
    \label{fig:new_result}
\end{figure}

Since data did not suggest any significant signal, we define a limit in the $(\delta m, \tau_{nn'})$ parameter space. The bound was computed as the $\chi^2(\delta m,\infty)+2^2$ contour line, which determines the 95\% C.L. exclusion region. The limit is displayed in Fig.\ \ref{fig:new_result} next to past results reported in UCN storage \cite{Ban_2007,Altarev_2009,berezhiani_2012,Berezhiani_2018_New,psi2021} and in regeneration experiments: MURMUR \cite{Stasser_2021}, STEREO \cite{stereo}, and SNS \cite{Broussard_2022}. The \textit{ratio channel} boundaries extracted from UCN storage experiments, which assumed mirror magnetic fields, were reinterpreted into the hidden matter scenario. The larger sensitivity reached with these storage experiments is explained first by the average free-flight time, which from MC simulations gives $(\bar{t}_f)\ssb{storage}/(\bar{t}_f)\ssb{beam}\sim 3$, and secondly by the fact that, since UCN are not stored in beam measurements, their average number of wall collision is smaller: $(\bar{n}\ssb{coll})\ssb{storage}/(\bar{n}\ssb{coll})\ssb{beam}\sim 150$.

Sources of systematic effects such as the detector response to magnetic fields and the background contamination were deeply studied \cite{saenz_phd}. Their influence on the ratio $R_{ABC}$ is negligible compared to the non-statistical fluctuations due to the reactor power variations. Moreover, deviations of the exclusion region were examined while modifying the MC input parameters. In particular, variations of the initial UCN velocity distribution and the probability for diffusive reflections resulted in less conservative limits: they increased the $\tau_{nn'}$ boundary over the whole mass splitting range by a $5\%$.

Reporting a single limit of $\tau_{nn'}$ for the entire scanned interval in this experiment is a cumbersome task if one wants to profit the maximum exclusion at each $\delta m$. Here, we chose the common conservative limit as the lowest valley of the 95\% contour line within the targeted interval:
\begin{equation}
    \tau_{nn'}> 1 \textrm{\,s for\,\,} |\delta m| \in [30 , 1143] \,\textrm{\SI{}{\micro\tesla}}\cdot \mu_n  \textrm{  (95\% C.L.)},
\end{equation}
or equivalently, in energy units
\begin{equation}
    \tau_{nn'}> 1 \textrm{\,s for\,\,} |\delta m| \in [2 , 69] \,\textrm{peV}  \textrm{  (95\% C.L.)}.
\end{equation}

The results presented in this work exclude $n-n'$ oscillations in a wide $\delta m$ range between UCN storage and regeneration measurements, with a sensitivity slightly lower than that of storage setups. While next generation neutron-sensitive neutrino setups and very-cold neutron experiments could still contribute to the search of hidden sectors through regeneration measurements for $\delta m$ up to \SI{1}{\micro\electronvolt}, neutron disappearance in UCN beams have the potential to probe the interval $\delta m \in [1 , 10^3]$ peV with improved sensitivities.

\begin{acknowledgments}
We would like to acknowledge the technical support by T. Brenner, D. Etasse, D. Goupillière, and N. Thiery during all the stages of the experiment. This work received financial support by the Swiss national science foundation grant number 200021\_169596.

\end{acknowledgments}

%\appendix

%\nocite{*}

\bibliography{apssamp}% Produces the bibliography via BibTeX.

\end{document}